\documentstyle[11pt,aaspp4]{article}

\newcommand{\vy}[2]{#1_{\rm\scriptscriptstyle #2}}

\begin{document}

\title{
	What is the redshift of gamma-ray burst 970508?
}
\author{
	Nahum Arav \& David W. Hogg
}
\affil{\sl
Theoretical Astrophysics, California Institute of Technology,\\
mail code 130-33, Pasadena, CA 91125, USA\\
{\tt arav, hogg@tapir.caltech.edu}
}

\begin{abstract}
A Bayesian likelihood analysis is used to constrain the redshift of
the optical transient associated with gamma-ray burst 970508, under
the assumption that the absorption at redshift 0.835 is not physically
associated with the transient.  The maximum-likelihood, expectation
value and 95-percent upper limit on the OT redshift are 0.835, 1.3 and
1.9 respectively.
\end{abstract}

\keywords{
methods:~statistical ---
quasars:~absorption lines ---
gamma~rays:~bursts ---
X-rays:~bursts
}

Gamma-ray bursts are among the most fascinating mysteries in
astrophysics, with their immense fluences, isotropic distribution on
the sky, and number-flux relation that is not consistent with
homogeneity.  The holy grail in the study of gamma-ray bursts is the
determination of their distances from the Earth; unfortunately, for
lack of solid evidence, distance estimates range from the outskirts of
our Solar System to the edge of the Universe.  The strongest
constraints on models are the intrinsic energies of the bursts, so
accurate distance determination is crucial.  A breakthrough occurred
this year with observations of an optical transient (OT), apparently
associated with an X-ray flare apparently associated with gamma-ray
burst 970508 (Bond 1997; Djorgovski et al 1997; Metzger et al 1997),
which show that the OT, and therefore probably the gamma-ray burst, is
at a cosmological distance.

The OT spectrum, taken near maximum light, shows a strong Mg~II
absorption system at redshift $z=0.835$ (Metzger et al 1997).  The
absorption system could be associated with the OT itself or else the
OT is further away and the gas cloud is between us and the OT.  If the
absorption is associated, the redshift of the OT is
$\vy{z}{OT}=0.835$.  There are several reasons to believe that the
absorption is associated: The absorption is strong (equivalent width
roughly 1.5\,\AA) and therefore the line-of-sight is not typical
(Steidel \& Sargent 1992).  Furthermore, the relative strengths of the
lines in the absorption suggests a line-of-sight through or at least
very near the center of a galaxy (Steidel, private communication),
which again makes the line-of-sight atypical.

However, if the absorption system is unrelated, $\vy{z}{OT}$ is not
known but can only be constrained.  In this {\sl Letter} we make the
constraint quantitative with a Bayesian likelihood analysis.  The
analysis is based on the known distribution of cosmological absorption
systems in high-redshift QSO spectra, which show either Mg~II (singly
ionized magnesium) or C~IV (triply ionized carbon) absorption.  The OT
is assumed to be at redshift $0.835<\vy{z}{OT}<2.3$ because no
Lyman-alpha ``forest'' lines are observed longward of $400$~nm
(Metzger et al 1997).  The probability of not having an absorption
system with $0.835<z<\vy{z}{OT}$ is computed and converted into a
likelihood function for $\vy{z}{OT}$ via Bayes's theorem.  This
procedure can easily be generalized for future observations by
replacing 0.835 with the highest redshift absorption system seen in
the spectrum of any cosmological source.  Our tacit assumption is that
in the OT spectrum there are no Mg~II or C~IV absorption systems at
redshifts $z>0.835$ down to the detection limit.  This is likely for
Mg~II systems the doublet is spectroscopically resolved so the systems
are easy to identify morphologically.  However, the wavelength
resolution of the spectrum is not sufficient to make morphological
identification of C~IV systems easy so it is possible that weak,
unidentified absorption lines in the spectrum could in fact be
high-redshift C~IV (Metzger, private communication).

By Bayes's theorem, the probability distribution function (pdf)
$p(\vy{z}{OT}|\mbox{no-abs})$ (probability per unit redshift) for the
OT redshift $\vy{z}{OT}$ given that there are no intervening
absorption systems at $z>0.835$ is given by
\begin{equation}
p(\vy{z}{OT}|\mbox{no-abs})=\frac{p(\vy{z}{OT})\,p(\mbox{no-abs}|\vy{z}{OT})}{p(\mbox{no-abs})}
\end{equation}
where $p(\vy{z}{OT})$ is a pdf summarizing our prior knowledge about
$\vy{z}{OT}$, $p(\mbox{no-abs}|\vy{z}{OT})$ is the probability of
observing no Mg~II or C~IV absorption systems given a particular value
of $\vy{z}{OT}$, and $p(\mbox{no-abs})$ is a normalization constant.
In this case the prior information is that the OT must have
$0.835<\vy{z}{OT}<2.3$ and here it is assumed that all redshifts in
this range are equally likely.  Perhaps a more natural choice is to
use a factor related to the comoving volume of the Universe but this
requires guessing the world model and the evolutionary behaviour of
the OT population, while influencing the final results very little.

The distribution of Mg~II absorption systems along QSO lines of sight
is well fit by a power law $dN/dz=N_0(1+z)^{\gamma}$ (number per unit
redshift per line of sight) with $\gamma=0.78$ (Steidel \& Sargent
1990) for absorption systems with rest frame equivalent width ${\rm
EW}>0.3$\,\AA.  The distribution of C~IV systems with ${\rm
EW}>0.3$\,\AA\ is well-fit with $\gamma=-0.90$ (Sargent, Boxenberg \&
Steidel 1988).  The amplitudes $N_0$ of these relations vary with
rest-frame EW limit $W_{\rm lim}$ as
\begin{equation}
N_0= N_{(0.3)}\,\exp \left(\frac{0.3\,{\rm \AA}-W_{\rm lim}}{W^{\ast}}\right)
\end{equation}
where parameters $(N_{(0.3)},\,W^{\ast})$ are (0.56,\,0.66\,\AA) for
Mg~II (Steidel \& Sargent 1990) and (10.9,\,0.41\,\AA) for C~IV
(Sargent et al 1988).  The estimated rest-frame equivalent-width
sensitivity of the OT spectrum (Metzger, private communication) is
plotted in Fig.~1, along with the differential number $dN/dz$ of
detectable systems per unit redshift, both as a function of redshift.

The mean expected number $m$ of independent absorption
systems along the line of sight from $z=0.835$ to $z=\vy{z}{OT}$ is
just the integral of the detectable $dN/dz$ over this range, where the
Mg~II $dN/dz$ is used at $z<1.52$ and is replaced by the C~IV $dN/dz$
at $z>1.52$ (when C~IV enters the spectroscopic window) because
essentially all Mg~II systems are also detected in C~IV (Steidel \&
Sargent 1990).  The probability $p(\mbox{no-abs}|\vy{z}{OT})$ of
getting no absorption systems is then simply $e^{-m}$.  This
probability is plotted in Fig.~2, along with the expectation value
$\vy{z}{OT}=1.3$ (integral of $z$ times $p(\vy{z}{OT}|\mbox{no-abs})$
over all redshifts) and the 95~percent confidence upper limit
$\vy{z}{OT}<1.8$ (point at which the integral of
$p(\vy{z}{OT}|\mbox{no-abs})$ is 0.95) marked.  The roughly 20~percent
uncertainties in the normalizations of the absorber distribution
$dN/dz$ lead to smaller than 5~percent uncertainties in the
expectation value and upper limit.  It is worthy of note that even
though the absorption system is not assumed to be associated, the
maximum likelihood redshift is very naturally $\vy{z}{OT}=0.835$.

\acknowledgements
We thank Mark Metzger and Chuck Steidel for information about the data
in advance of publication, and Roger Blandford, Shri Kulkarni and Wal
Sargent for useful comments.  Support by the NSF is gratefully
acknowledged.

\newpage
\figcaption[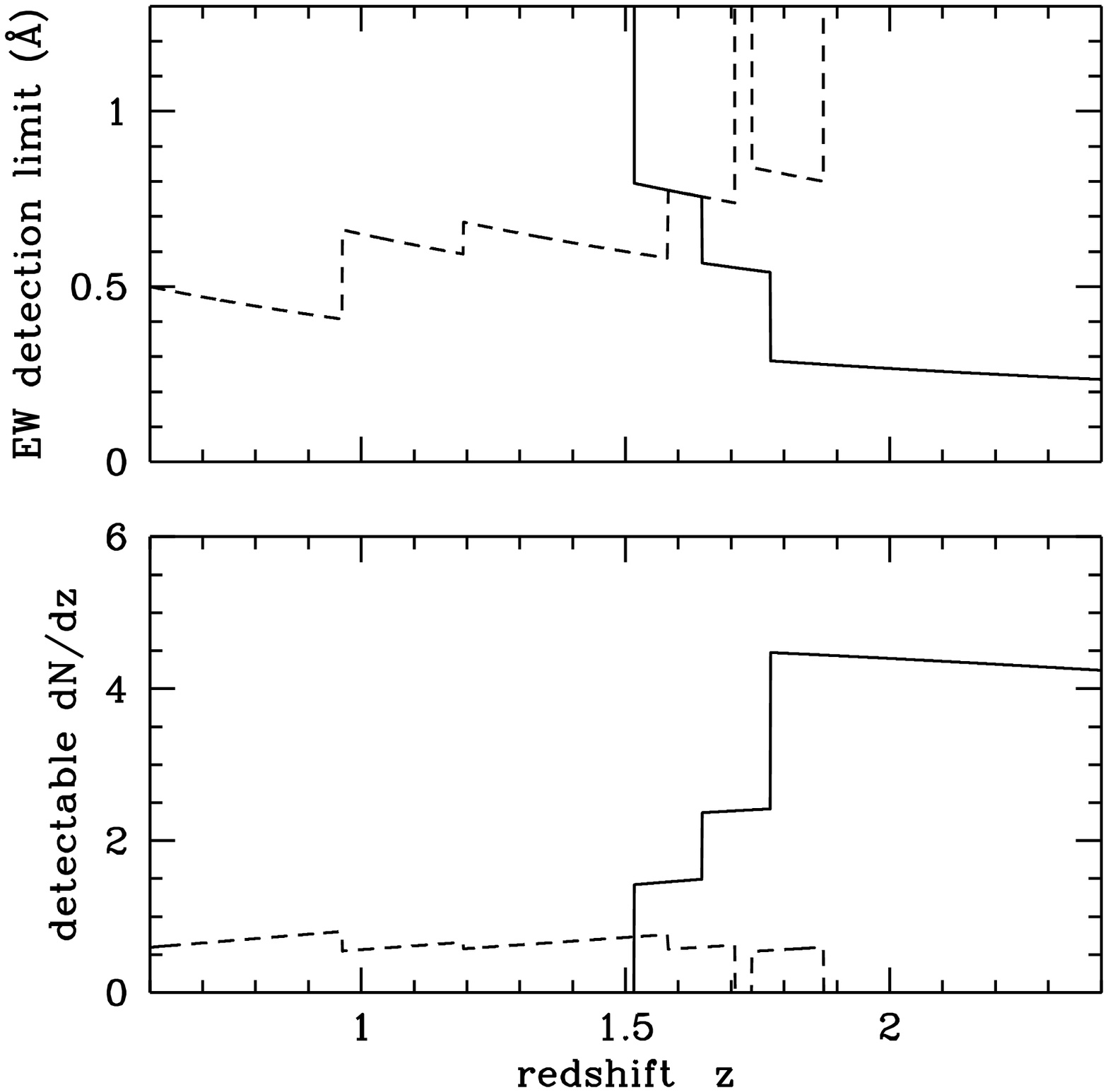]{
{\sl (Top)}~The rest-frame equivalent width detection limit (observed
detection limit divided by $1+z$) of the OT spectrum, as reported by
Metzger (private communication), appropriate for Mg~II (dashed curve)
and C~IV (solid).  The spike in the Mg~II limit at $z\approx 1.7$ is
caused by a sky line, Mg~II leaves the spectroscopic window at
$z\approx 1.9$ and C~IV enters at $z\approx 1.5$.  {\sl (Bottom)}~The
differential number $dN/dz$ of absorption systems expected per line of
sight for Mg~II (dashed, Steidel \& Sargent 1990) and C~IV (solid,
Sargent et al 1988), down to the detection limit plotted in the {\sl
(Top)} panel.}

\figcaption[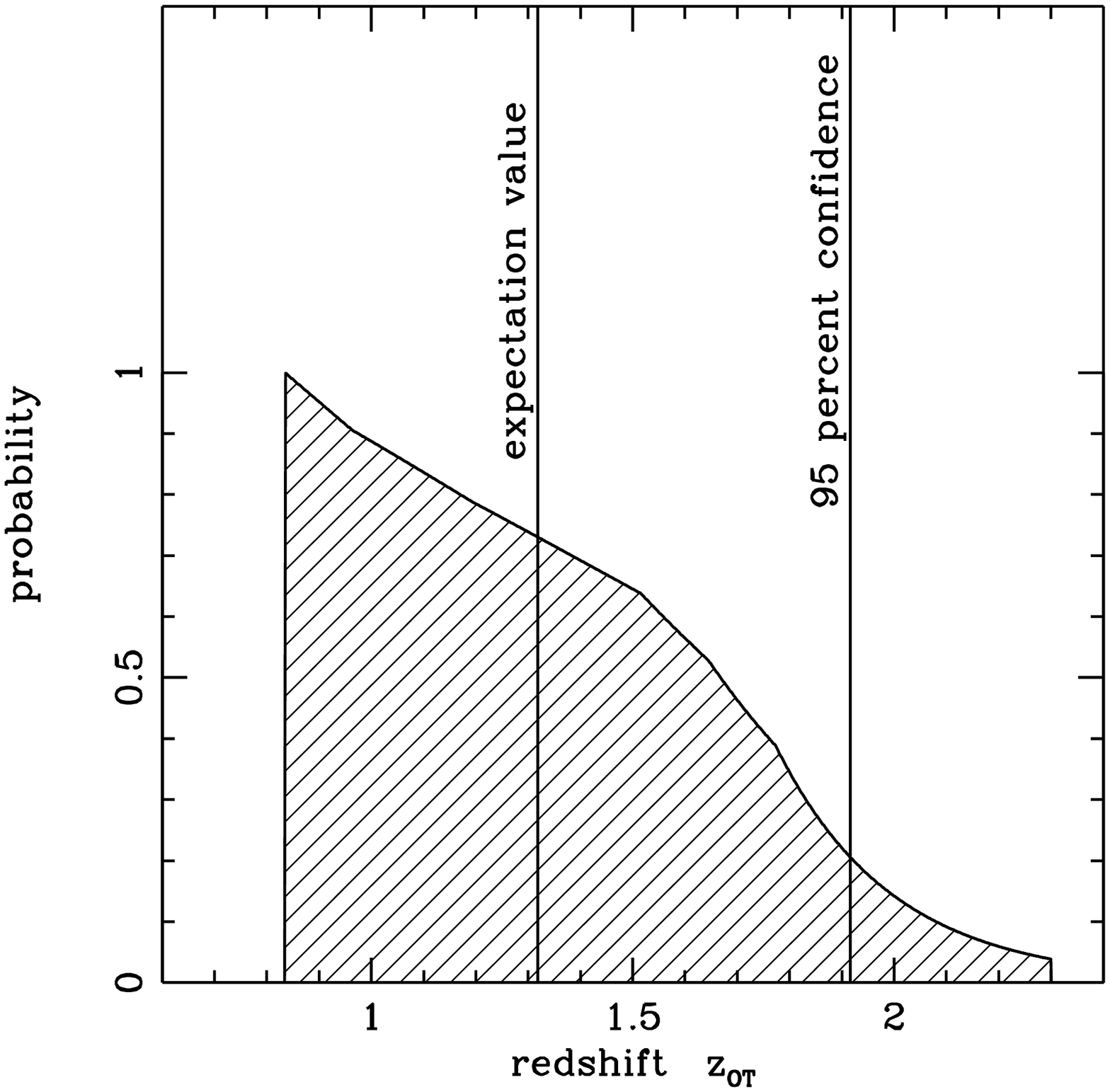]{
The probability that a source at redshift $\vy{z}{OT}$ will show no
Mg~II or C~IV absorption systems at $z>0.835$ as a function of
redshift, times the prior probability that the source is in the range
$0.835<\vy{z}{OT}<2.3$.  This is proportional (by Bayes's theorem) to
the pdf (probability per unit redshift) for the redshift $\vy{z}{OT}$
of the OT given that it shows no intervening absorption systems at
redshifts $z>0.835$.  The steepening at $\vy{z}{OT}\approx 1.6$
results from the great increase in the expected number of observable
absorbers when C~IV enters the spectroscopically observed range.  The
expectation value for the redshift and the 95~percent confidence level
upper limit are marked with vertical bars.}

\setcounter{figure}{0}
\begin{figure}
\plotone{detectable.eps}
\caption{~}
\end{figure}
\begin{figure}
\plotone{probdist.eps}
\caption{~}
\end{figure}

\end{document}